\documentclass[aps,twocolumn,superscriptaddress,groupedaddress,nofootinbib]{revtex4-1}
\usepackage{graphicx}
\usepackage{bm}       
\usepackage{amssymb}
\usepackage[toc,page]{appendix}
\usepackage{braket}
\newcommand{\del}{\partial}

\usepackage{amsmath}

\begin{document}

\title{
Gravity theory with a dark extra dimension}

\author{Sandipan Sengupta}
\email{sandipan@phy.iitkgp.ac.in}
\affiliation{Department of Physics and Centre for Theoretical Studies, Indian Institute of Technology Kharagpur, Kharagpur-721302, INDIA}

\begin{abstract}

We set up a vacuum theory of gravity with an extra dimension of vanishing proper length. The most general solution to the field equations are presented. This formulation is free of Kaluza-Klein modes and does not allow the propagation of gravitons along the invisible fifth direction. Apart from a vacuum energy and radiation, the associated emergent theory exhibits a nonpropagating vector-tensor multiplet which has no analogue in standard Einstein gravity. It is naturally inert, obeys a bounded equation of state and has coupling properties radically different from ordinary matter. Based on these distinctive features, we propose that this geometric multiplet could supercede the hypothetical `dark matter'. As further evidence in support of this possibility, we show that the galactic rotation curves are predicted to be asymptotically flat.


\end{abstract}

\maketitle

\section{Introduction}

The physics of extra dimensions, apart from just being
a fascinating idea, has often provided uncanny insights
into some of the outstanding problems in general relativity and particle physics. This idea probably owes the first real impetus to Kaluza through his attempt
to unify gravitational and electromagnetic interactions,
along with Klein who proposed that the extra spatial dimension is curled up in a small circle to justify why it might
not be observable \cite{kaluza,klein,albert}. Further conceptual progress have 
emerged through the application of this framework in efforts to understand the hierarchy between the weak and Planck
scale \cite{dvali,rs}. However, the fact that there are no signs of
Kaluza-Klein excitations or other possible signatures of small or large \cite{ant} extra
dimensions as yet makes it worthwhile to explore alternative ideas.

Here we introduce a different formulation of extra dimen-
sions altogether. This is based on a dynamical theory of
vacuum gravity where the fifth dimension has a vanishing
metrical length, and thus is unobservable in principle. The
appropriate five-dimensional action principle, which is
shown to admit such spacetime solutions, is defined through
a first-order formulation.

First, we obtain the most general solution to the resulting
field equations. The detailed structure of the associated
emergent theory in four dimensions is also unraveled. This
is characterized by an invertible metric and additional
fields originating from the nondynamical connection components. The phenomenological prospects are rich and
intriguing. This theory does not exhibit any propagating emergent fields other than the four metric, or any excitation such as the Kaluza-Klein modes.

The effective stress tensor contains a pure trace, which represents a classical vacuum energy. The equation of state of the remaining (nonpropagating) field multiplet as an ideal fluid is found to be bounded both above and below. In the four dimensional effective action resulting from the five dimensional vacuum theory, these fields do not generate any interaction other than gravitational. Also, its coupling to any additional matter that could possibly be introduced in four dimensions is shown to have a natural suppression through the four-dimensional Planck scale. 

Following the general structure of this theory as well as the distinctive features of the nonpropagating field content, we show that it could supercede the notion of `dark matter', which has long been conjectured to be one of the missing elements of standard Einsteinian gravity \cite{zwicky,smith,rubin,roberts,persic,batt,peebles}. This is what we propose here. 

Furthermore, we set up a model of the galactic halo based on this extra dimensional framework. A strong support to our proposal above is provided by the behaviour of the resulting rotation curves \cite{rubin,roberts}, whose large distance profile is different from that expected due to ordinary matter. The emergent fields making up the (nonlumionous) halo naturally explain its stability against nongravitational interactions in general. We conclude our analysis with the discussion of the propagation of graviton within this theory.

\section{The fundamental theory}
While introducing the Lagrangian formulation, we assume that the independent fields are the vielbein $\hat{e}_\mu^I(x)$ and super-connection $\hat{w}_\mu^{~IJ}(x)$ ($\mu\equiv (t,x,y,z,w)$, $I\equiv (0,1,2,3,4)$). 
The five-dimensional theory in vacuum is defined by the Hilbert-Palatini functional:
\begin{eqnarray*}
{\cal L}(\hat{e},\hat{w})=\frac{1}{L^3}\epsilon^{\mu\nu\alpha\beta\rho} \epsilon_{IJKLM}\hat{e}_{\mu}^{I}\hat{e}_{\nu}^{J}\hat{e}_{\alpha}^{K} \hat{R}_{\beta\rho}^{~~LM}(\hat{w}),
\end{eqnarray*}
where $L$  is the (fundamental) five-dimensional Planck-length and $\hat{R}_{\beta\rho}^{~~LM}(\hat{w})=\del_{[\beta} \hat{w}_{\rho]}^{~LM}+\hat{w}_{[\beta}^{~LK}\hat{w}_{\rho]K}^{~~~M}$ is the field-strength. The internal metric is defined as $\eta_{IJ}=[-1,1,1,1,\sigma]$, where $\sigma=\pm 1$. 
The field equations are obtained directly from the variation of the above:
\begin{subequations}
\begin{align}
\epsilon^{\mu\nu\alpha\beta\rho} \epsilon_{IJKLM}\hat{e}_{\mu}^{I}\hat{e}_{\nu}^{J} \hat{D}_{\alpha}(\hat{w})\hat{e}_{\beta}^{K}=0,\label{eom1}\\
\epsilon^{\mu\nu\alpha\beta\rho} \epsilon_{IJKLM}\hat{e}_{\mu}^{I} \hat{e}_{\nu}^{J} \hat{R}_{\alpha\beta}^{~KL}(\hat{w})=0\label{eom2}
\end{align}
\end{subequations}
Here we have defined $\hat{D}_\mu$ as the gauge-covariant derivative with respect to the super-connection $\hat{w}_\mu^{~IJ}$.

In general, the field equations (\ref{eom1}) and (\ref{eom2}) admit solutions which could have either vanishing or nonvanishing determinant of the vielbein. To explore the consequence of an extra dimension of zero metrical length in gravity theory (in vacuum), here we look to find the most general solution to the above equations where the vielbein has a zero eigenvalue. This zero may be chosen to lie along a particular direction ($v$):
\begin{eqnarray}
\hat{e}_v^I=0.
\end{eqnarray}
This implies:
\begin{eqnarray*}
\hat{e}_\mu^I =
\left[\begin{array}{cc}
\hat{e}_a^i\equiv e_a^i & 0 \\
0 & 0 \\
\end{array}\right]
\end{eqnarray*}
Note that $e_a^i$ may be viewed as the emergent tetrad fields (invertible) that define the geometry of the four-dimensional spacetime. We denote the inverse of these tetrads as $e^a_i$ ($\neq \hat{e}^a_i$, since $\hat{e}^\mu_I$ do not exist):
\begin{eqnarray*}
e_a^i e^b_i=\delta_a^b,~e_a^i e^a_j=\delta _j^i.
\end{eqnarray*}
The corresponding $4$-metric is defined as $g_{ab}=e_a^i e_{bi}$.

\subsection{General solution to the connection equations of motion}
The connection equations (\ref{eom1}) are linear in the connection fields. These decouple into two sets:
\begin{eqnarray}\label{e1}
\epsilon^{abcd} \epsilon_{IJKLM}e_{a}^{I}e_{b}^{J} \hat{D}_{v}e_{c}^{K}=0,\nonumber\\
\epsilon^{abcd} \epsilon_{IJKLM}e_{a}^{I}e_{b}^{J} \hat{D}_{c}e_{d}^{K}=0.
\end{eqnarray}
Let us obtain the most general solution to these as discussed below.

The first set above imply:
\begin{eqnarray}
\hat{D}_v e_a^i=0,~\hat{D}_v e_a^4=0.
\end{eqnarray}
These may be solved for the connection components $\hat{w}_v^{~IJ}$, which are manifestly trivial upto a pure gauge:
\begin{eqnarray}\label{gauge}
\hat{w}_v^{~ij}=-e^{cj}\del_v e_c^i,~\hat{w}_v^{~4i}=0.
\end{eqnarray}
Note that eqn.(\ref{gauge}) implies that  
the metric $g_{ab}$ is independent of the fifth coordinate $v$:
$\del_{v}g_{ab}=\del_v (e_a^i e_{bi})=e_{ai} \hat{D}_v e_b^i+(a\leftrightarrow b)=0$. Hence, any $v$-dependence in the tetrad must be  purely a gauge artefact, and could be transformed away by an appropriate gauge choice. This is equivalent to choosing the unphysical connection components to vanish, without any loss of generality:
\begin{eqnarray}\label{gauge1}
\hat{w}_v^{~ij}=0,
\end{eqnarray}
 implying $\del_v e_c^k=0$.

In the second set, the $M\neq 4,L\neq 4$ component leads to six equations:
\begin{eqnarray}
e^{c[i} \hat{w}_c^{~j]4}=0 \Rightarrow \hat{w}_a^{~4i}=M^{ij}e_{aj}
\end{eqnarray}
where the arbitrary spacetime field $M^{ij}=M^{ji}$ represents a $4\times 4$ matrix.
In other words, these six equations fix the antisymmetric components of $\hat{w}_a^{~4i}$ to be zero, whereas the remaining ten are left arbitrary.
Next, the $M=4$ component of the same set implies:
\begin{eqnarray}\label{w2}
e^d_k \hat{D}_{[c}e_{d]}^k=0,
\end{eqnarray}
This has the general solution:
\begin{eqnarray}
\hat{w}_a^{~ij}=\bar{w}_a^{~ij}(e)+K_a^{~ij},
\end{eqnarray}
where $\bar{w}_a^{~ij}(e)=\frac{1}{2}[e^b_i\del_{[a}e_{b]}^j
-e^b_j\del_{[a}e_{b]}^i -  e_a^l e^b_i e^c_j
\del_{[b}e_{c]}^l]$ are the torsionless connection components ($D_{[a}(\bar{w})e_{b]}^i=0$) and the contortion tensor $K_a^{~ij}=-K_a^{~ji}$ is constrained as:
\begin{eqnarray}\label{kconstraint}
e^a_j K_a^{~ij}\equiv K_a^{~ia}=0.
\end{eqnarray}
We may parametrize the remaining twenty independent components of the contortion in terms of an axial vector field $L^i$ and a tensor field $N^{ijk}=-N^{ikj}$ with $N_i^{~ik}=0,~\epsilon_{ijkl}N^{jkl}=0$ ($20=4+16$):
\begin{eqnarray}\label{K}
K_a^{~ij}=\epsilon^{ijkl}e_{ak} L_l+2e_{al} N^{lij}.
\end{eqnarray}  
Note that since torsion is assumed to be even under parity, $L_i$ and $N_{ijk}$ must have mutually opposite parity.


\subsection{General solution to vielbein equations of motion}

The vielbein equations (\ref{eom2}) may be divided into the following sets:
\begin{eqnarray}\label{e2}
&& \epsilon^{abcd} \epsilon_{IJKLM}\hat{e}_{a}^{I} \hat{e}_{b}^{J} \hat{R}_{cv}^{~~KL}(w)=0,\nonumber\\
&& \epsilon^{abcd} \epsilon_{IJKLM}\hat{e}_{a}^{J} \hat{e}_{b}^{K} \hat{R}_{cd}^{~~LM}(w)=0.
\end{eqnarray}
In the following, we find the general solutions for each set separately for the $I\neq 4$ and $I=4$ components, respectively.

Among the first set, for $I\neq 4$ we have:
\begin{eqnarray}
\hat{R}_{av}^{~~i4}=0 \implies \hat{D}_v M^{kl}=0,
\end{eqnarray}
Using eq.(\ref{gauge}), this implies that $M^{kl}$ (and hence $\hat{w}_a^{~4i}$) are $v$-independent.
The $I=4$ component, on the other hand, is identically satisfied. 

In the second set, the $I\neq 4$ component leads to further constraints:
\begin{eqnarray}\label{meq}
e^a_k \hat{R}_{ab}^{~~k4}= 0&\implies & \left[\delta^a_b \delta_{kl}-e^a_k e_{bl} \right] D_a M^{kl}=0
\end{eqnarray}
Note that the covariant derivative $D_a$ is defined with respect to the torsionfree connection $\bar{w}_a^{~ij}(e)$.
Finally, the $I=4$ component of the second set in eq.(\ref{e2}) reads:
\begin{eqnarray}\label{e3}
e^a_i e^b_j \hat{R}_{ab}^{~~ij}= 0.
\end{eqnarray}

\subsection{Independence of the fifth coordinate}
The fact that the fields $e_a^i$ and $\hat{w}_a^{~5i}$ (and hence $M_{ij}$) are independent of the fifth coordinate $v$ has already been elucidated earlier. To find the dependence of the remaining connection components on $v$, let us first note the identity given by:
\begin{eqnarray}
\epsilon_{IJKLM} \epsilon^{\mu\nu\alpha\beta\rho}\hat{e}_{\mu}^{I} \hat{e}_{\nu}^{J} \hat{D}_\alpha \hat{R}_{\beta\rho}^{~~KL}(\hat{w})=0.
\end{eqnarray}
The $M=4$ component of the above implies:
\begin{eqnarray}
&&\del_v \left[\epsilon^{abcd} \epsilon_{jklm}e_{a}^{j} e_{b}^{k} \hat{R}_{cd}^{~~lm}(w)\right]\nonumber\\
&& +2 \epsilon^{abcd} \epsilon_{jklm}e_{a}^{j} e_{b}^{k} \hat{D}_c \hat{R}_{dv}^{~~lm}(w)=0
\end{eqnarray}
However, the term within the bracket in the first piece vanishes due to the field equation (\ref{e3}). Simplifying the second term further,
 we obtain:
 \begin{eqnarray}
 e^{a}_{[i} e^{b}_{j]} \hat{D}_a\left[\del_v \hat{w}^{~ij}\right]=0=\del_{v}\left[e^a_i e^b_j K_{[a}^{~ik}K_{b]k}^{~~~j}\right]
 \end{eqnarray}
 This  shows that the only contortion-dependent piece which appears in the equation of motion (\ref{e3}) is $v$ independent. 
 Using the generic decomposition (\ref{K}) of the contortion, this becomes equivalent to the constraint:
  \begin{eqnarray}
   3 L^i \del_v L_i+ N^{ijk} \del_v N_{ijk}=0
\end{eqnarray}  
However, since the fields $L_i$ and $N_{ijk}$ are linearly independent, the above relation implies:
\begin{eqnarray}
\del_v L_i=0=\del_v N_{ijk}.
\end{eqnarray}
Thus, all the basic fields are independent of the fifth coordinate, a feature that emerges naturally within this framework\footnote{This may be contrasted to the case of Kaluza-Klein theory, where one requires to impose additional assumptions (i.e. the cylinder condition) in order to obtain such a property.}.


\section{Structure of the emergent theory}
Let us note that the equation of motion (\ref{e3}) fixes the scalar component of the field-strength to zero. This implies the following general solution:
\begin{eqnarray}\label{ricci}
\hat{R}_{ab}^{~~ij}=\hat{t}_{ab}^{~~ij},
\end{eqnarray}
where $\hat{t}_{ab}^{~~ij}$ is an arbitrary tensor field satisfying $e^a_i e^b_j \hat{t}_{ab}^{~~ij}=0$ and having the same symmetry properties as the field-strength tensor. Using the identity: $\hat{R}_{ab}^{~~ij}=\bar{R}_{ab}^{~~ij}(\bar{w})+D_{[a}(\bar{w})K_{b]}^{~ij}+K_{[a}^{~ik}K_{b]k}^{~~~j}-\sigma M^{i}_{~k} M^{j}_{~l} e_{[a}^k e_{b]}^l$
where $\bar{R}_{ab}^{~~ij}(\bar{w})=\del_{[a} \bar{w}_{b]}^{~ij}+ \bar{w}_{[a}^{~ik}\bar{w}_{b]k}^{~~~j}$ is the torsionfree part of the field-strength, we may rewrite the above solution as:
\begin{eqnarray}\label{R0}
\bar{R}_{ab}^{~~ij}(\bar{w})&=&\hat{t}_{ab}^{~~ij}-D_{[a}(\bar{w})K_{b]}^{~ij}-K_{[a}^{~ik}K_{b]k}^{~~~j}\nonumber\\
&+&\sigma M^{i}_{~k} M^{j}_{~l} e_{[a}^k e_{b]}^l
\end{eqnarray}
In terms of the decomposition $K_a^{~ij}\equiv (L_i,N_{ijk})$ as in eq.(\ref{K}), the third piece at the right hand side above becomes:
\begin{eqnarray*}
&&K_{[a}^{~ik}K_{b]k}^{~~~j}=L^k L_k e_{[a}^i e_{b]}^j-L^{[i} L_{[a} e_{b]}^{j]}\nonumber\\
&&+2 e_{b}^m e_{a}^n [N_m^{~ki}N^{~~~j}_{nk}-N^{i}_{~mk} N^{jk}_{~~~n}]\nonumber\\
&&-2[e_{am} e_{bn} \left(2N^{mi}_{~~~k}N^{nkj}+N_k^{~in} N^{jmk}+N_k^{~jm} N^{nik}\right)\nonumber\\
&&-e_{[a}^p e_{b]}^q L^n(\epsilon^{j}_{~pnk}N_{q}^{~ki}-\epsilon^{i}_{~pnk}N_{q}^{~kj})]
\end{eqnarray*}
The above identity results in a simpler expression for the field-strength obtained in eq.(\ref{R0}):
 \begin{eqnarray}\label{riemann1}
&&\bar{R}_{ab}^{~~ij}(\bar{w})=\bar{t}_{ab}^{~~ij}-L^k L_k e_{[a}^i e_{b]}^j+L^{[i} L_{[a} e_{b]}^{j]}\nonumber\\
&&-2 e_{b}^m e_{a}^n [N_m^{~ki}N^{~~~j}_{nk}-N^{i}_{~mk} N^{jk}_{~~~n}]
+\sigma e_{[a}^k e_{b]}^l M^{i}_{~k} M^{j}_{~l} 
\end{eqnarray}
where the terms having the same property as the tensor $\hat{t}_{ab}^{~~ij}$ have been absorbed away through its redefinition: 
\begin{eqnarray*}
\bar{t}_{ab}^{~~ij}&=&\hat{t}_{ab}^{~~ij}-\left[D_{[a}(\bar{w})K_{b]}^{~ij}\right]\\
&-&2 e_{am} e_{bn} \left[2N^{mi}_{~~~k}N^{nkj}+N_k^{~in} N^{jmk}+N_k^{~jm} N^{nik}\right]\\
&+& 2 e_{[a}^p e_{b]}^q L_n \left[\epsilon^{jpnk}N_{qk}^{~~i}-\epsilon^{ipnk}N_{qk}^{~~j}\right]
\end{eqnarray*}

From the field-strength (\ref{riemann1}), it is straightforward to find the (torsionless) Riemann tensor associated with the general solution discussed above:
\begin{eqnarray}
\bar{R}_{abcd}&=&-L^e L_e g_{c[a}g_{b]d}+L_c L_{[a} g_{b]d}-L_d L_{[a}g_{b]c}\nonumber\\
&+&\sigma M_{c[a} M_{b]d}-2\left[N_{bck}N_{ad}^{~~~k}-N_{cbk}N_{da}^{~~~k}\right],
\end{eqnarray}
where, we have lowered (raised) the internal indices using the tetrad (inverse tetrad) fields: $L_a\equiv L^i e_{ai},~M_{ab}\equiv M^{ij} e_{ai}e_{bj},~N_{ab}^{~~~k}\equiv N^{ijk}e_{ai}e_{bj}$. Note that this tensor has the usual symmetries: $\bar{R}_{abcd}=\bar{R}_{cdab}=-\bar{R}_{bacd}=-\bar{R}_{abdc}$. However, the standard identity $\bar{R}_{[abc]d}=0$ for an Einsteinian Riemann tensor is satisfied provided: 
\begin{eqnarray}
N^{ijk}=N^{jik}.
\end{eqnarray}
However, this, along with its original symmetry properties eq.({\ref{K}}), leads to:
\begin{eqnarray}
N_{ab}^{~~k}=0=N^{ijk}.
\end{eqnarray}
Thus, the equation of motion and the symmetry properties of the Riemann tensor imply that the contortion can only be made up of the axial four-vector $L^i$.

\subsection{Effective Einstein equations}

Based on the results above, the emergent Einstein equations read:
\begin{eqnarray}\label{EE1}
&&\bar{R}_{ab}-\frac{1}{2}g_{ab}\bar{R}=\bar{t}_{ab}+[2L_a L_b+L_c L^c g_{ab}]\nonumber\\
&&+\sigma[(M^c_{~c} M_{ab}-M_a^{~c} M_{cb})-\frac{1}{2} \left(M^c_{~c} M^d_{~d}-M^{cd}M_{cd}\right) g_{ab}],\nonumber\\
~
\end{eqnarray}
the Ricci tensor and scalar being defined as $\bar{R}_{ab}=\bar{R}_{ac}^{~~ij} e_j^c e_{bi},~\bar{R}=g^{ab}\bar{R}_{ab}$.
This governs the emergent gravity theory in four dimensions. Note that the effective stress-energy tensor at the right hand side is purely geometric, where the symmetric tensor $\bar{t}_{ab}=\bar{t}_{ac}^{~~ij} e_j^c e_{bi}$ is traceless and the fields $L^a,M^{ab}$ originate from the nonpropagating components of the connection $\hat{w}_\mu^{~IJ}$, exhibiting no second order time derivatives. However, these fields are not completely arbitrary, since their first space and time derivatives are restricted through the Bianchi identity that they must satisfy:
\begin{eqnarray}
&&\nabla_b (\bar{t}^{ab}+[2L^a L^b+L_c L^c g^{ab}]\nonumber\\
&&+\sigma[(M^c_{~c} M^{ab}-M^a_{~c} M^{cb})-\frac{1}{2} \left(M^c_{~c} M^d_{~d}-M^{cd}M_{cd}\right) g^{ab}])\nonumber\\
&&=0
\end{eqnarray}
where $\nabla_a$ is the covariant derivative defined with respect to the Christoffel symbols $\Gamma_{ab}^{~~c}(g)$ made up of the emergent (invertible) metric $g_{ab}$.

In the general solution (\ref{EE1}) resulting from the basic five-dimensional theory, the first piece $\bar{t}_{ab}$ represents radiation. Before going on to analyze the properties of the remaining field multiplet, let us note that the field $M_{ij}$ admits an orthogonal decomposition into a scalar (trace) $\chi$ and a tracefree part ($10=1+9$) in general:
\begin{eqnarray}\label{M}
M_{ij}=\frac{1}{4}\chi\eta_{ij}+\sqrt{2}S_{ij},~\eta_{ij}S^{ij}=0.
\end{eqnarray}
In terms of these, the effective Einstein equations finally simplify to:
 \begin{eqnarray}\label{EE2}
 \bar{R}_{ab}-\frac{1}{2}g_{ab}\bar{R}&=&t_{ab}-\frac{3\sigma}{16}\chi^2 g_{ab}+[2L_a L_b+L_c L^c g_{ab}]\nonumber\\
&-&\sigma\left[2 S_a^{~c}S_{cb}-S_{cd}S^{cd}g_{ab}\right] ,
 \end{eqnarray}
 where the traceless piece has been redefined as $t_{ab}=\left[\bar{t}_{ab}+\frac{1}{2}MS_{ab}\right]$.
 The second term quadratic in $\chi$ essentially represents a variable vacuum energy. 

\subsection{A special case: Cosmological constant}

We may consider a scenario where the radiation piece and the rest are separately conserved:
\begin{eqnarray}
\nabla_b \bar{t}^{ab}=0&=&\nabla_b ([2L^a L^b+L_c L^c g^{ab}]\nonumber\\
&-&\sigma[2S^a_{~c} S^{cb}-S^{cd}S_{cd} g^{ab}]-\frac{3\sigma}{16}\chi^2 g^{ab})\nonumber\\
&=&2L^c \left[\delta_c^b \nabla_a L^a+
\nabla_c L^b+g_{ab}\del_a L^c \right]\nonumber\\
&&+S^{cd}[g^{ab}\nabla_{a}S_{cd}-\delta_d^b \nabla_a S^a_c-\delta_d^a \nabla_a S^b_c]\nonumber\\
&&-\frac{3\sigma}{16}g^{ab}\chi\del_a \chi
\end{eqnarray}
Under the assumption that the fields $\chi$, $L^a$ and $S^{ab}$ are linearly independent, this implies:
\begin{eqnarray}
&& \del_a \chi=0,\nonumber\\
&&\nabla_a L_b+\nabla_b L_a=0,\nonumber\\
&& g^{bd}\nabla_a S^{ac} +g^{ad}\nabla_a S^{bc}-g^{ab}\nabla_a S^{cd}=0.
\end{eqnarray}
In other words, this sector of the theory inherits a cosmological constant ($\chi\equiv$ spacetime const.), whose origin is geometric.

\subsection{Four dimensional effective action}
Here we provide an  effective four dimensional action formulation equivalent to the one discussed above. For simplicity, we shall assume a trivial radiation field ($t_{ab}=0$). 

Due to the vanishing proper length of the extra dimension, a naive dimensional reduction by integrating over the fifth direction does not work here. Rather, the appropriate action is generated by the four dimensional Hilbert-Palatini term, along with a Lagrange multiplier field to implement the associated constraints:
\begin{eqnarray}\label{eff}
&&S_{eff}(e_a^i,w_a^{~ij},\zeta^a)=\frac{1}{2l_P^2}\int d^4 x~e[e^a_i e^b_j R_{ab}^{~~ij}(w)
\nonumber\\
&&~+~  2\zeta^a e^b_i D_{[a}(w)e_{b]}^i- e^a_i e^b_j J_{ab}^{~~ij}(M)],
\end{eqnarray}
where the source is defined as $J_{ab}^{~~ij}(M)=\sigma M_{[a}^i M_{b]}^j$ and the covariant derivative is defined by the general connection field  (with contortion).

The variation of this action with respect to $\zeta^a$ yields the following equations of motion:
\begin{eqnarray*}
e^b_i D_{[a}(w)e_{b]}^i\approx 0,
\end{eqnarray*}
which are equivalent to (\ref{w2}), obtained earlier from the five-dimensional Lagrangian. The weak equality `$\approx$' implies that these equations of motion are to be implemented only after all the variations have been performed. Next, the connection equations imply:
\begin{eqnarray*}
\zeta^a =\frac{1}{3} g^{ab}\left[e^c_k D_{[b}e_{c]}^k\right]\approx 0.
\end{eqnarray*}
Finally, a variation with respect to the tetrad, upon implementing the above two field equations, leads to the same Einstein equation as in (\ref{EE1}).

As reflected by the reduced action (\ref{eff}), our formulation is fundamentally different from variants of modified theories of gravity, such as $f(R)$, $f(T)$ or conformal gravity. Whereas such theories generically exhibit higher curvature modifications and/or propagating torsion, the fields $L_i,M_{ij}$ here have no kinetic terms and do not propagate. From the four-dimensional perspective, these are equivalent to source fields which as a whole are associated with a conserved current (due to the Bianchi identity). 

The framework presented here is essentially different from any extra dimensional formulation where the fifth dimension has a nontrivial proper length, such as the Kaluza-Klein approach. To recall, the emergent four dimensional fields here do not resemble ordinary propagating matter, and the four dimensional Planck length is not generated from the five-dimensional length scale through a dimensional reduction. In addition, since the fifth coordinate has no period, there are no tower of higher eigenmodes with discrete momenta to be looked for in the collider experiments. 
 


\section{Distinctive properties of the emergent multiplet and implications}

\subsection{An ideal fluid description}
The equations of state of the radiation piece $t_{ab}$ and the vacuum energy $\chi$ are well-known ($\omega=\frac{1}{3}$ and $-1$, respectively).
 Let us rewrite the geometric contribution (\ref{EE2}) to the stress tensor in a perfect fluid form:
\begin{eqnarray}
&&[2L_a L_b+L_c L^c g_{ab}]-\sigma\left[2 S_a^{~c}S_{cb}-S_{cd}S^{cd}g_{ab}\right]\nonumber\\
&&=(\rho+P)u_a u_b +P g_{ab}
\end{eqnarray}
where $u^a$ is the four-velocity of the fluid. Since the fields $L_a,S_{ab}$ are independent of the metric, the coefficients of $g_{ab}$ at both sides above must be equal.
As a result, the effective density and pressure could be solved as:
\begin{eqnarray}\label{P}
&& \rho=-3L_a L^a+\sigma S_{ab}S^{ab}= -3L_i L^i+\sigma S_{ij}S^{ij}\nonumber\\
&& P=L_a L^a+\sigma S_{ab}S^{ab}=L_i L^i+\sigma S_{ij}S^{ij}
\end{eqnarray}
There are a number of important features, as elucidated by the above equation:

(a) The axial vector field $L_i$ corresponds to negative pressure $\omega_L=-\frac{1}{3}$, whereas the symmetric traceless tensor $S_{ij}$ resembles a stiff fluid with $\omega_s=1$;

(b) Assuming an equation of state of the form $P=\omega\rho$ for the emergent composite ($L^i,S^{ij}$), the non-negativity of the corresponding energy densities  $\rho_L$ and $\rho_S$ implies that the 
equation of state  is bounded both above and below:
\begin{equation}\label{omega}
-\frac{1}{3}\leq \omega\leq 1;
\end{equation}  

(c) This prediction could have important implications in the context of cosmological evolution. Generally, at the earliest stages the dominant fluid component should be $S_{ab}$ whereas the axial field $L_i$ dominates the latest stages (till the vacuum energy takes over). A  pressureless composite is expected to characterize an intermediate phase. This corresponds to the constraint $L_i L^i=-\sigma S_{ij}S^{ij}$, with a relative density $\frac{\rho_S}{\rho_L}=\frac{1}{3}$. 

\subsection{Couplings}

The emergent fields $L_i$ ($L_a$) and $S_{ij}$ ($S_{ab}$) are dimensionful. This is because they originate from the (five dimensional) super-connection $\hat{w}_\mu^{~IJ}\equiv (\hat{w}_\mu^{~ij},\hat{w}_\mu^{~i4})$, which has the dimension of inverse length, whereas the pentads $\hat{e}_\mu^I$ are dimensionless.
Let us define this length scale as $l$. 
This, also being the scale that defines the emergent cosmological constant (when $\chi=constant$), must generally be much larger than the four-dimensional Planck length ($l_P $).


Note that the gravitational coupling constant of the emergent stress-tensor in eq.(\ref{EE2}) is independent of the Newton's constant $G$. 
To compare with the stress-energy tensor associated with ordinary matter, we may rewrite this equation in terms of the standard coupling as:
\begin{eqnarray*}
\bar{R}_{ab}-\frac{1}{2}g_{ab}\bar{R}=8\pi G T^{(eff)}_{ab}
\end{eqnarray*} 
In the above, $T^{(eff)}_{ab}$ has the dimension $[T^{(eff)}_{ab}]=\frac{1}{Gl^2}$, and has no inherent mass scale. This is in stark contrast to normal matter with $[T^{(m)}_{ab}]=\frac{GM}{l^3}$, where $M$ defines the mass of the field producing the energy-momentum tensor. In other words, whereas the density of ordinary matter behaves as $\rho_{(m)}\sim \frac{M}{l^3}$ at some distance scale $l$, the effective energy density of  these emergent fields goes as $\rho_{(eff)}\sim \frac{1}{Gl^2}$. The latter becomes dominant at a large distance in a region devoid of matter, owing to its slower fall-off with distance.
Hence, we are led to a remarkable conclusion, that is, the emergent field content $(L_a,S_{ab})$ in eq.(\ref{EE2}) does not resemble ordinary matter.

The absence of any mass scale in the effective gravitational coupling above also means that these fields do not have a particulate nature. In other words, these cannot exchange energy through collisions with matter particles and hence would be noninteracting in general. Note that these are also non-propagating.
Nevertheless, it is possible in principle to introduce additional couplings of these dimensionful fields to four dimensional propagating matter (i.e., associated with at least two derivatives in the Lagrangian) in the emergent theory. For instance, the simplest possible nontrivial couplings of these to some scalar matter $\phi$ would be given by $L^{(1)}_\phi\sim l_P\sqrt{-g} S^{ab} \del_a \phi \del_b \phi,~L^{(2)}_\phi\sim l_P^2\sqrt{-g} L^aL^b \del_a \phi \del_b \phi$ etc. However, such couplings are manifestly suppressed by positive powers of the four dimensional Planck length $l_P$, which is the only fundamental length scale in the emergent theory. This shows that even if any additional matter is introduced, their couplings to these geometric fields would naturally be too weak to be relevant.

\subsection{A proposed resolution to the `dark matter' problem}

The discussion above leads us to the main application of the theoretical framework presented here. Let us recall some of the notable features that have been unravelled till now:

(i) The nonprogating components of the super-connection reflected by the fields $(L_{i},S_{ij})$ do not behave as ordinary matter, as the respective couplings are very different;

(ii) Due to its noninvertibility, the 5-metric $\hat{g}_{\mu\nu}$ cannot be coupled directly to (bosonic) matter fields in five dimensions. Therefore, the fundamental five-dimensional theory does not generate any coupling of the fields $L_i,S_{ij}$ to matter in the effective four-dimensional description;

(iii) Any additional matter coupling that could possibly be introduced at the four dimensional level to these fields are Planck-scale suppressed. Hence, such couplings are also too weak to be relevant; 

(iv) Their coupling to gravity has a slower fall-off compared to ordinary matter and hence is expected to dominate at large scales where there is no luminous matter;

(v) This coupling is independent of any field mass scale, implying that the associated fields cannot be interpreted as ordinary particulate matter which may collide and dissipate energies. Hence, these fields are inert in general;

(vi) The predicted equation of state of this composite is bounded. In particular, this bound contains pressureless fluid as an intermediate point, which is believed to represent the effective nature of the conjectured `dark matter' at the current epoch.

Based on these observations, we propose that the emergent field content ($L_i,S_{ij}$) should supercede the notion of `dark matter' altogether. Note that the inertness, which is a nontrivial feature but is essential for anything which could be conceived to play the role of some hypothetical `invisible matter', seems to emerge naturally owing to the zero proper length of the extra dimension.

We shall now explore the implications of our general formulation for galactic rotation curves. This is particularly important, since the flatness of the rotation profiles of spiral galaxies as confirmed in the seventies \cite{rubin,roberts,persic,batt} has been one of the main motives behind the `dark matter' conjecture.

\section{Prediction of flat rotation curves}
We have already found how effective density of the geometric field composite $(L_a,S_{ab})$ varies with distance. 
Assuming that these are the essential constituents of the galactic halo which is spherical symmetric, this implies: $\rho_{eff}(r)\sim \frac{1}{r^2}$, or,  $M_{eff}(r)\sim r$ for the corresponding effective `mass'. For any test particle moving in a circular orbit under the influence of this `mass' within the radius $r$, the equality of the gravitational pull with the centripetal force leads to a the constant circular velocity. It seems remarkable that the emergent theory constructed earlier naturally predicts that the rotation curves should be flat at some sufficiently large distance. 
This may be contrasted with standard Einsteinian gravity where coupled to normal matter through $G$ such a scenario is simply untenable. The inherent inertness of these fields, as already explained earlier, also implies that the halo would effectively be invisible to particulate matter. Hence, it would be stable against interactions in general.

Based on these observations, we now set up a geometric model of the halo and look for possible new insights into the general relativistic solutions corresponding to flat rotation curves. For simplicity, we shall ignore the (subdominant) effects of baryonic mass or vacuum energy and also of the radiation term $t_{ab}$.


We begin with the standard approach of describing the outer region of a galaxy by a spherically symmetric static metric \cite{nuc}:
\begin{eqnarray*}
ds^2=-e^{\mu(r)}dt^2+e^{\lambda(r)} dr^2+r^2(d\theta^2+\sin^2 \theta d\phi^2)
\end{eqnarray*}
The circular velocity of the test particles at a radial distance $r$ reads \cite{nuc}:
\begin{eqnarray*}\label{vc}
v^2(r)=e^{-\mu}r^2 \left(\frac{d\phi}{dt}\right)^2=\frac{1}{2}r\mu'(r).
\end{eqnarray*}
With the above metric, the three independent diagonal components of the effective Einstein's equations (\ref{EE1}) lead to (assuming isotropic pressure):
\begin{eqnarray}\label{EE3}
e^\lambda \rho &=& \frac{\lambda'}{r}-\frac{1}{r^2}(1-e^{\lambda})\nonumber\\
e^\lambda P &=& \frac{\mu'}{r}+\frac{1}{r^2}(1-e^{\lambda})\nonumber\\
e^\lambda P &=& \mu''+\left(\frac{\mu'}{2}+\frac{1}{r}\right)\left(\mu'-\lambda'\right)
\end{eqnarray}
Note that the effective density $\rho$ and and the pressure $P$, defined in (\ref{P}), refers to the geometric field composite only and not to any ordinary matter. 
In view of the spherical symmetry of this model, we assume the fields $(L^i,S^{ij})$ to depend on the radial coordinate $r$ only. 

\subsection{Newtonian limit:}

The Newtonian solution to the above set of equations corresponding to a constant circular velocity is
 well-known \cite{nuc}.
In the limit of a slowly varying $\lambda(r)$ ($|\lambda'(r)|\ll |\mu'(r)|$) and small pressure, eqs.(\ref{EE3}) can be solved for the two metric functions as:
\begin{eqnarray}
\lambda(r)=A, ~\mu(r)= (A-1)\ln r+B,
\end{eqnarray}
where $A, B$ are constants. From eq.(\ref{vc}), we see that these correspond to a constant circular velocity $v$ with $2v^2=A-1$. The axial and non-axial densities become, respectively:
\begin{eqnarray}
\rho_L= \frac{3}{4r^2}(2v^2-v^4),~\rho_S= \frac{1}{4r^2}(2v^2+3v^4)
\end{eqnarray}
This describes the present phase of the halo fluid during the course of cosmological evolution, as is suggested by current observations.

\subsection{Non-Newtonian limit:}

However, there also exists a non-Newtonian particular solution to the equations (\ref{EE2}) with constant terminal velocity, which does not really have a theoretical explanation as to what is the field that could produce it. 
This is obtained in the limit of large effective pressure: $P\gg \frac{\mu'(r)}{r}$:
\begin{eqnarray}
\rho_L(r)&=& \frac{3C}{2}\left[\frac{2-v^2}{1-v^2}\right]r^{\frac{2v^2}{1-v^2}},~\rho_S(r)= \frac{C}{2}\left[\frac{v^2}{1-v^2}\right]r^{\frac{2v^2}{1-v^2}},\nonumber\\
P(r)&=&-C r^{\frac{2v^2}{1-v^2}},~\lambda(r)=-ln\left[1-C r^{\frac{2}{1-v^2}}\right],
\end{eqnarray}
where  $C>0$ is an integration constant.
The associated state parameter is given by $\omega=-\left[\frac{1-v^2}{3-v^2}\right]\approx -\frac{1}{3}$. 
Our framework, however, does have a field component which exhibits precisely this equation of state, namely, the pseudovector field $L_i$. Evidently, this limit corresponds to a state of the halo fluid that is composed almost entirely of the axial 4-vector $L^i$:
\begin{eqnarray}
\frac{\rho_L}{\rho}=\frac{3(2-v^2)}{2(3-v^2)}\approx 1,~\frac{\rho_S}{\rho}=\frac{v^2}{2(3-v^2)}\approx 0.
\end{eqnarray} 
This phase is expected to dominate the latest stage of the cosmological evolution, since the density contribution due to $L_i$ should have the slowest fall-off owing to its extreme equation of state.

To summarize, the non-Newtonian features of the flat rotation curve geometry has a natural explanation in terms of the vector contortion $L_i$, while the Newtonian features have an interpretation in terms of a combined effect of $L_i$ and the symmetric traceless tensor $S_{ij}$. 

\subsection{Numerical estimates}
Note that the axial density $\rho_L$ could in fact be determined from observational data. For instance, using the radial variation of the halo mass function $M_H(r)$ given by $
\del_r M_H(r)=4\pi r^2 \rho (r)$ (ignoring the baryonic mass within the galactic disc) along with $P=\omega \rho$ and eq.(\ref{P}), we obtain:
\begin{eqnarray}
\rho_L(r)=\frac{3(1-\omega)}{2}\rho=\frac{3(1-\omega)}{8\pi r^2}\del_r M_H(r).
\end{eqnarray}
Since $\del_r M_H(r)$ is an observable that may be determined from the study
of gravitational lensing within or outside the halo \cite{narayan},
the axial density could be found from the above, given the equation of state. Alternatively, 
$\rho_L$ may be determined provided the total effective density of the halo fluid $\rho$ is known. A comparison between the two values so obtained could be one way to check the consistency of this geometric model of the galactic halo. For instance, using the known numbers $v\approx 238$ km/s, $\rho \approx 0.4$ Gev $cm^{-3}$ for the Milky way halo \cite{sofue1}, the Newtonian solution corresponds to $w \approx 3\times 10^{-6}$ (at the present epoch). From this the densities of the two independent components are estimated to be:
\begin{eqnarray}\label{halo}
\rho_{L}\approx 0.3 ~Gev~ cm^{-3},~ \rho_{S}\approx 0.1 ~Gev ~cm^{-3}.
\end{eqnarray}

\subsection{Unification of three length-scales}

Based on the fact that the emergent vacuum energy is defined by the trace of $M^{ij}$, which are some of the superconnection components, we may find the typical order of magnitude of the length scale $l$ associated with the emergent fields. Based on the current observational data related to the cosmological constant $\Lambda$,  we obtain:\\
\begin{eqnarray}
\chi^{-1}\sim \frac{1}{\sqrt{\Lambda}}\sim 10^{26} m
\end{eqnarray}
An intriguing question is, could there be a characteristic length scale in galatic physics which corresponds to this order of magnitude. Empirically, we do know that a possible length scale is given by the typical galactic accelerations: $a_0\sim 2\times 10^{-10} ms^{-2}$, which  for $c=1$ reduces to: $a_0\sim 10^{-26} ~ m^{-1}$. This is precisely the inverse of the number obtained above. Hence, from the  perspective of the five dimensional theory, the galactic acceleration scale is not fundamental, but is derived from that of the superconnection, and is also related to the vacuum energy scale. 

\section{Propagation of graviton}

Finally, we come to the question as to how does the graviton mode propagate in this theory. Note that both the unperturbed as well as perturbed 5-metrics should be defined to be solutions the first-order equations of motion in this formulation, and hence both must be degenerate (and hence they satisfy the first order field equations presented earlier)\footnote{This is exactly analogous to the case of invertible tetrads where both the unperturbed and perturbed metrics are required to satisfy Einstein's field equations.}. The corresponding 4-metrics, however, may be assumed to be nondegenerate in general. Even though one could in principle consider the possibility of having a perturbed 5-metric that is invertible, that would correspond to a different theory, something beyond the scope or context of the present article. 

Recalling our earlier discussion, the unperturbed 5-metric is given by:
\begin{eqnarray*}
\hat{\bar{g}}_{\mu\nu} =
\left[\begin{array}{cc}
\bar{g}_{ab} & 0 \\
0 & 0 \\
\end{array}\right]
\end{eqnarray*}
The most general perturbation on this is represented by:
\begin{eqnarray*}
ds^2=[g_{ab}-\sigma h^2 A_a A_b]dx^a dx^b+ \sigma h^2 [dv+A_a dx^a]^2
\end{eqnarray*}
where $\sigma=\pm 1,~g_{ab}=\bar{g}_{ab}+h_{ab},~g_{av}= g_{va}=h^2 A_a,~g_{vv}=h^2$.
The associated pentad fields read:
\begin{eqnarray*}
\hat{e}_\mu^I =
\left[\begin{array}{cc}
e_a^i & h A_a \\
0 & h \\
\end{array}\right]
\end{eqnarray*}
Since the determinant of the pentad $\hat{e}=eh$ ($e$ is the determinant of the 4-metric) must vanish, we have $h=0$. Hence, in general, the perturbed 5-metric is necessarily of the following form:
\begin{eqnarray*}
\hat{g}_{\mu\nu} =
\left[\begin{array}{cc}
\bar{g}_{ab}+h_{ab} & 0 \\
0 & 0 \\
\end{array}\right]
\end{eqnarray*}
Note that the 4-fluctuation $h_{ab}$ are subject to the effective Einstein equation (\ref{EE1})  along with the four dimensional Bianchi identities. This leads to an equation with second order time derivatives on $h_{ab}$, exactly as in the case of (linearized) Einstein's equation with a source. Evidently, in the general set of solutions found earlier, there is no other second order equation involving any of the remaining fields. 

To emphasize, the (two transverse traceless polarizations of) graviton propagates only in four dimensions. This feature reflects another important difference from the Kaluza-Klein based extra dimensional formulations \cite{dvali,rs,ant}.

\section{Conclusions}

We have introduced a first order theory of vacuum gravity with an extra dimension of vanishing proper length, and have explored the dynamical consequences. 
The emergent four dimensional theory, apart from providing a geometric realization of vacuum energy, exhibits a nonpropagating vector-tensor multiplet. It is shown to have characteristics radically different from ordinary matter. This analysis provides a concrete basis to our subsequent proposal that this emergent field content, which has no analogue within standard Einsteinian theory, leads to a potential resolution to the `dark matter' problem. This is one of the main imports of the fundamental theory introduced here.

Remarkably, a model of the galactic halo based on this extra dimensional formulation predicts that the rotation curves should be flat at large distances where
there is no effect of visible matter. The characteristic inertness of the vector-tensor multiplet also provides a natural explanation to the stability of the halo.

In the emergent multiplet, the (axial) vector component has an equation of state $\omega=-\frac{1}{3}$ which provides a possible explanation of the non-Newtonian features of the flat rotation curve solutions. The Newtonian features, on the other hand, are explained in terms of a pressureless composite which requires the vector as well as the tensor component $S_{ij}$ which has a stiff equation of state $\omega=1$. Both these kinds of fields could have their own implications. For instance, these fields could play a role in the late and early stages of a cosmological evolution, respectively. To mention a few special contexts in isolation, the possible existence of stiff matter in relativistic theories has been discussed earlier by Zeldovich \cite{zel}, whereas the equation of state $\rho+3P=0$ has been found to emerge in the context of cosmic strings \cite{gott}.
It seems intriguing though that such nonstandard equations of state could also be a property of fields of purely geometric origin, as elucidated here. This begs for a deeper understanding.

The dynamical embedding of one or more extra dimension of
zero length within an action formulation and its possible implications, as explored here, have not been studied before. It is worthwhile to note that the geometric and topological properties (e.g. geodesic completeness, space or time orientability etc.)  of the associated solutions are fundamentally distinct 
from degenerate spacetime solutions in four dimensional gravity, some of which have been discussed earlier in various contexts \cite{tseytlin,horowitz,kaul,kaul2,san,san1,gera}. 

As an infrared modification of gravity, our formulation is manifestly different from the modified theories in general (such as higher curvature gravity or MOND or conformal gravity  etc. \cite{milgrom,mannheim}).
Also, the  additional dimension of zero proper length here does not inherit the defining features of the earlier frameworks based on compact or large or warped dimensions \cite{kaluza,klein,rs,dvali}. To emphasize, this theory is free of the Kaluza-Klein tower of particles, and the graviton propagates only in four dimensions. It seems remarkable that the independence of the effective four dimensional theory of the fifth coordinate emerges as a natural consequence of the field equations, without requiring any additional assumption as such. Furthermore, this theory is not a limiting case of a Kaluza-Klein theory, where the length of the variable fifth dimension could be made to approach zero. In fact, such a limit is singular, which is not a surprise since the two theories correspond to two different sets of equations of motion.

Even though there exists no metrical scale originating from the extra dimension  (unlike the
Kaluza-Klein or Randall-Sundrum or ADD case), the emergent fields do exhibit a characteristic length different from the four dimensional Planck length. This is also the scale characterising the vacuum energy (cosmological constant being a special case) on one hand and the typical galactic accelerations on the other, thus providing an explicit connection between two large numbers appearing in nature. One wonders whether this fact has a deeper relevance.

It is worth mentioning that a universal acceleration (length) scale  has some times been invoked to define the working formula designed to fit the flat rotation curves, even though it is not clear from current observations if such a universal scale does exist. However, within any such phenomenologically motivated formulation (e.g. MOND), the origin of a new scale has no explanation. From our viewpoint here, such a large unit of length rather is derived from the fundamental scale of the (super-) connection fields of the basic five dimensional theory. 

In general, this scale could have nontrivial imports in the weak or Planck scale physics. The generic implications of this framework on cosmological dynamics, too, seem rich in prospects and possible surprises.

\begin{acknowledgments}
I am indebted to F. Barbero for a critical reading of the manuscript and to A. Ghosh and U. Sarkar for numerous helpful discussions. I also thank S. Bharadwaj, K. Dutta, R. Kaul and P. Majumder for their comments. The support of (in part) the SERB, DST through the grant ECR/2016/000027 and of (in part) the ISIRD grant RAQ  are gratefully acknowledged.
\end{acknowledgments}

\end{document}